\documentclass[preprint]{aastex}

\newcommand{\per}{$^{-1}$}

\shorttitle{The Molecular Gas Core of HH 212}
\shortauthors{Wiseman et al.}

\begin{document}


\title{The Flattened, Rotating Molecular Gas Core of Protostellar Jet HH~212}




\author{Jennifer Wiseman\altaffilmark{1}}
\affil{Department of Physics and Astronomy, The Johns Hopkins University,
Baltimore, MD 21218}

\author{Alwyn Wootten}
\affil{National Radio Astronomy Observatory, Charlottesville, VA 22903}

\author{Hans Zinnecker and Mark McCaughrean}
\affil{Astrophysikalisches Institut, Potsdam, Germany D-14492}




\altaffiltext{1}{Hubble Fellow}


\begin{abstract}

The recently discovered protostellar jet known as HH212 is beautifully
symmetric, with a series of paired shock knots and bow shocks on
either side of the exciting source region, IRAS 05413-0104 (Zinnecker
{{\it{et~al.}}} 1998).  We present VLA ammonia maps of the IRAS
05413-0104 molecular gas envelope in which the protostellar jet source
is embedded.  We find that the envelope, with mass of 0.2 M$_{\odot}$
detected by the interferometer, is flattened perpendicular to the jet
axis with a FWHM diameter of 12000 AU and an axis ratio of 2:1, as
seen in NH$_3$ (1,1) emission.  There is a velocity gradient of about
4-5 km sec$^{-1}$ pc$^{-1}$ across the flattened disk-like core,
suggestive of rotation around an axis aligned with the jet.
Flux-weighted mean velocities increase smoothly with radius with a
roughly constant velocity gradient.  In young (Class 0) systems such
as HH212, a significant amount of material is still distributed in a
large surrounding envelope, and thus the observable kinematics of the
system may reflect the less centrally condensed, youthful state of the
source and obscuration of central dynamics.  The angular momentum of
this envelope material may be released from infalling gas through
rotation in the HH212 jet, as recent observations suggest (Davis
{\em{et~al.~}}2000).  A blue-shifted wisp or bowl of emitting gas
appears to be swept up along the blue side of the outflow, possibly
lining the cavity of a wider angle wind around the more collimated
shock jet axis.  Our ammonia (2,2)/(1,1) ratio map indicates that this
very cold core is heated to 14 Kelvin degrees in a centrally condensed
area surrounding the jet source.  This edge-on core and jet system
appears to be young and deeply embedded.  This environment, however,
is apparently not disrupting the pristine symmetry and collimation of
the jet.

\end{abstract}




\section{Introduction}

The youngest protostars are often so deeply embedded in their parental
cloud cores that shocks driven by their jets cannot be easily traced 
in visible shock
line emission but are bright in near-infrared molecular hydrogen
shock lines.  Such is the case for the recently discovered highly
collimated jet HH 212, near the Horsehead Nebula region of Orion
(Zinnecker, McCaughrean, \& Rayner 1998).

HH 212 is remarkable because of its high collimation and symmetric
pairs of bow shocks and shock knots 
on either side of the driving source (Figure 1).
Theories of protostellar jet structure often involve a heavy influence
of the surrounding medium, with impacts creating shocked globules
along the jet, sometimes even redirecting it.  But the symmetry of HH
212 suggests that some pulsing mechanism from the source itself is
responsible for the nearly periodic bow shock pairs.  Somehow the
symmetry is maintained even though the jet is moving through a dense
gas core.   In the innermost region, water masers have been found in
motion along the outflow (Claussen et~al. 1998).

The inferred position of the jet exciting source coincides with IRAS
point source 05413-0104, which was detected at 25, 60, and 100
microns; it is still too cold and embedded to be detected strongly at
12 microns (see Beichman et al. 1986).  It is surrounded by a cold
(11.5 K), compact ammonia core out of which it presumably formed
(Wouterloot, Walmsley, \& Henkel 1988).  The spectral energy
distribution climbs to longer wavelengths.  It is detected also at 1.3
mm, with a submillimeter to bolometric luminosity of about 2\% (Chini
et al. 1997). It has narrow NH$_3$ linewidths compared to emission
found at the position of other maser sources.  These characteristics
suggest that the source is very young.

HH212 is classified as a Class 0 system, the youngest class of
protostars with much material to be accreted still in the surrounding
molecular gas envelope (Andr\'e, Ward-Thompson, \& Barsony 2000).
Class 0 sources have a high rate of submillimeter to bolometric
luminosity (over 0.5\%) and high rates of infall and outflow (Andr\'e,
Ward-Thompson, \& Barsony 1993, Adams, Lada, \& Shu 1987).  The
morphology and kinematics of the surrounding gas at this early stage
of the formation of a star are crucial to understand as an important
link between prestellar gas cores and the resulting young T Tauri
star/disk systems.  Our VLA ammonia observations were conducted as a
study of the gas density, temperature, and velocity distributions in
this pristine, highly embedded young source.

\section{Observations}

Observations were carried out with the
Very Large Array (VLA) of the National Radio Astronomy
Observatory\footnote{The National Radio Astronomy Observatory is a
facility of the National Science Foundation operated under cooperative
agreement by Associated Universities, Inc.}.  The ammonia (J,K) =
(1,1) and (2,2) inversion transitions were observed simultaneously,
at rest frequencies of 23.694495 GHz and 23.722733 GHz,
respectively.  The width of the antenna primary beam response was
approximately 2' FWHP.  The array was in its compact, ``D'' configuration.
The synthesized beamsize was $8.4'' \times 8.8''$ FWHM, with natural
weighting and a 20 k$\lambda$ taper applied to the UV data.
The bandwidth was 3.125 MHz and the channel separation was 24.4
kHz, corresponding to 0.31 km sec$^{-1}$.

The absolute flux calibrator was 3C286, with a calculated flux density
of 2.41 Jy at 1.3 cm.  The bandpass calibrator was 3C84, and a phase
closure solution was applied based on interleaved observations of the
phase calibrator B0605-085 (1950).
The ``dirty'' beam was deconvolved from the maps using the ``Imager''
algorithm in the NRAO AIPS image processing software.

\section{A Flattened, Rotating, and Heated Gas Envelope}

The integrated intensity of the NH$_3$ (1,1) line is displayed in the
contours of Figure 1, overlaid on the 2.12 $\micron$ image of HH 212
from Zinnecker, McCaughrean, \& Rayner (1998).  The gas core 
is flattened with an axis ratio of about 2:1.  The major axis has
a FWHM extent of 29'', corresponding to 12000 AU at an assumed distance
of 400 pc, with full extent
of over 15000 AU;  it is centered on the jet source, and
it lies perpendicular to the jet which has a position angle of
about 25 degrees east of north and only 4$\deg$ from the plane
of the sky (Claussen et al. 1998).

A velocity gradient is apparent across the flattened gas core.  In
Figure 2a the integrated intensity (moment 0) is again plotted in
contours, while the underlying shades represent the (moment 1)
velocity field.  Radial velocities change smoothly from blue-shifted
gas in the northwest side to redshifted gas in the southeast; the
large velocity gradient of about 4-5 km sec\per pc\per is virtually
constant.  The direction of the gradient within the central 
FWHM intensity region of the core is perpendicular
to the jet axis, with the nodes of constant velocity roughly parallel
to the jet. 

Along the eastern side of the core, a small amount of blueshifted
material is seen to extend to the northeast alongside the northern,
blueshifted side of the jet.  To a lesser extent, the same effect is
also seen to the south with a small protrusion of redshifted gas
extending in the same direction as the redshifted side of the jet.
Molecular gas is possibly being swept along by the outflow.  There is
even a hint, at least in the north, that molecular gas is found in a
``bowl'' surrounding the jet axis, possibly outlining a cavity of a
wider-angle flow surrounding the collimated shock emission of the jet
(see, e.g., Shu et~al. 1995, Shang, Shu,\& Glassgold 1998).

Ammonia inversion lines are a convenient probe of both optical depth
and temperature.  Most ammonia inversion transitions are composed of
multiple hyperfine components. The relative intensities of central and
satellite component blends of the (1,1) transition can be used to
calculate the optical depth.  For the main component and total
transition of the (1,1) line, we find $\tau$ values of 0.8 and 1.6,
respectively, at the core center.  Using the procedures and
assumptions described in Ho \& Townes 1983 (see also Wiseman \& Ho
1998), we also estimate an NH$_3$ column density of $1.6 \times
10^{14}$ cm$^{-2}$ and a total volume density $n$ of $1 \times 10^5$
cm$^{-3}$.  The mass of the core gas detected with the interferometer
is approximately 0.2 M$_{\odot}$,
assuming an NH$_3$/H$_2$ abundance ratio of $2 \times 10^{-8}$ (Harju,
Walmsley, \& Wouterloot 1993).  Within the large uncertainties of this
abundance ratio, the mass is the same as that needed to
gravitationally bind the rotating system.

The ratio of intensity of the NH$_3$ (2,2) and (1,1) transitions
reflects level populations and thus serves as a thermometer for
rotational temperature within cold molecular gas cores.  In Figure 2b,
the contours again show the integrated intensity of (1,1) emission.
The shades show the (2,2)/(1,1) ratio, indicating the relatively
warmer regions.  It is apparent that this is indeed a cold core of
gas, with the (2,2) emission showing up only in the central region of
the core.  The (2,2)/(1,1) ratio peaks near the position of the
embedded protostellar source, and extends slightly to the northwest.
The peak rotational temperature here is 14 K, as calculated from the
line ratio (Ho \& Townes 1983).  In colder regimes such as this, the
kinetic temperature is approximately equal to the NH$_3$ rotational
temperature (Danby et al. 1988, Kuiper 1994).  The young protostar is
evidently still in the center of the gas core, and we may be seeing
here the warm gas {\em{heated}} by the protostar and the jet as they
interact with the environment.

\section{Discussion}

\subsection{Rotation in Protostellar Systems}

Envelopes of Class 0 protostars carry much (often most) of the mass of
the system and can extend to diameters of several thousands of AU.
Interferometric maps of such systems tend to show flattened disks
perpendicular to the axis of the associated jet (Myers, Evans, \&
Ohashi 2000).  The flattened molecular gas envelope of the HH 212 jet
source has a smooth and roughly constant velocity gradient
perpendicular to the jet axis (Figure~2a,~2c). 

Turbulence in cloud cores can produce velocity gradients that mimic
rotation (Burkert \& Bodenheimer 2000 (BB)).  However, the gradient we
find in HH 212 is faster than the most likely effect
produced in the BB turbulence models.  It should be noted that the BB
models were based on more massive cores, so proper scaling of the
models would be required to better assess the possibility of a
turbulent component to the observed velocity gradient in HH 212.  But
it is also true that the HH 212 gradient extends systematically across
the long axis of the core, and perpendicular to the jet, for over 5
synthesized beamwidths.  Thus the combined evidence suggests that the
gradient is more likely dominated by rotation.

The more evolved T Tauri systems often show Keplerian rotation
signatures in the less plentiful envelope gas, with the radial
velocities increasing and diverging toward the central source
($V\propto R^{-1/2}$).  An early report of Keplerian rotation in a
young stellar disk was made for $^{13}$CO observations of HL~Tau
(Sargent \& Beckwith 1987, 1991), though in this source the kinematics
proved to be much more complicated, with evidence for infall (Hayashi
et al.  1993) and outflow (Cabrit et al. 1996).  The strongest
Keplerian velocity curves are seen in circumstellar disks of a few
hundred AU in Class II T~Tauri systems without much envelope material
left, thus having little confusion from large scale infall or outflow
motions (Mundy, Looney, \& Welch 2000, Dutrey et al. 1998).  Keplerian
rotation has been detected in the gas disks of GM~Aur (Dutrey et
al. 1998) and DM~Tau (Guilloteau \& Dutrey 1998), for example, as well
as in multiple systems such as UY~Aur and the beautiful
ring of GG~Tau (Duvert et al. 1998, Guilloteau, Dutrey, \& Simon 1999).

Several Class 0 systems such as HH 212 are in dense gas cores that
appear to have a more rigid rotation signature (Myers, Evans, \&
Ohashi 2000); this is consistent with their character of having much
of the system mass still distributed in the gas envelope around the
embedded accreting protostar(s).  Indeed, the 0.2 M$_\odot$ we detect
with the VLA in the HH 212 core is 10 times the circumstellar gas mass
found in some T Tauri systems (e.g. $\le 0.01$M$_{\odot}$ for UY Aur
(Duvert et al. 1998)).  In such early systems, a gradient in the
intensity-weighted mean velocity that is constant rather than
diverging toward the center can indicate that either a) the
non-Keplerian rotation curve is reflecting a mass distribution that is
not yet dominated by a central condensation, or b) the system is in
Keplerian rotation, but high optical depths result in velocity maps
that are dominated by motion in the outer layers along the line of
sight toward the center of the core.  Models of expected channel map
emission can be used to test for Keplerian rotation (e.g. Koerner,
Sargent, \& Beckwith 1993, Dutrey, Guilloteau, \& Simon 1994).  Higher
spatial and spectral resolution maps would be needed to fully model
the HH 212 envelope, although the slightly broader line widths we
observe toward the center of the core are possible evidence for some
Keplerian component to the motion.  In any case, it now appears 
that {\em{the observed kinematics of the
protostellar envelope can signify the developmental stage of the
embedded YSO}}.  More high resolution kinematical studies are needed
to confirm this pattern.  The large envelopes such as we observe in
HH~212 may feed mass directly to the embedded protostar (if at the
earliest stage) or to the inner circumstellar accretion disk just
beginning to form.  It is important to note that this inner disk ($<
100$ AU and thus unobservable in our NH$_3$ maps) probably rotates
with near-Keplerian velocities (Stahler 2000).

 Many rotating cores of these large sizes ($\ge$12000 AU) do
not appear to conserve specific angular momentum, defined as V$_{rot}$
$\times$ R$_{rot}$, where V$_{rot}$ is the velocity of rotation at
radius R$_{rot}$ (Myers, Evans, \& Ohashi 2000).  Our results show
that, with a specific angular momentum of 0.012 km s$^{-1}$ pc at a
radius of 6000 AU, the HH 212 gas envelope would fit well with other
cores of similar size in a plot of Ohashi et al. (1997) showing
specific angular momentum falling with decreasing source radius.
However, Ohashi et al. showed that there is a transition at about this
same size scale; smaller dense core regions {\em{do}} show conserved
specific angular momentum in regions of both massive infall and
rotationally supported disks.  Thus this would serve as the size scale
for dynamical collapse, where angular momentum is conserved.  The HH
212 core seems to be in fast, bound rotation at the size scale of this
transition.

Since the velocity gradient and the long axis of the envelope are both
perpendicular to the jet axis, it is suggestive that rotation may
contribute to the observed flattening.  We calculate rotational energy
of the system to reach up to about 3-10\% of each of the thermal,
turbulent, and gravitational energy (Harju et al. 1993, Myers 1983).
Preferential contraction along an axis perpendicular to the disk
plane, along possible magnetic field lines, could also produce
flattened morphology.  Another intriguing possibility is that the
outflow itself contributes to the morphology by clearing out dense gas
along the flow.  Indeed, the blue-shifted wisp of gas extending to the
northeast alongside the (northward) blue-shifted side of the jet
(Figure 2a) seems to form part of a ``bowl'' around the jet and the
associated CO outflow recently studied by Lee et al. (2000); part of
the blue lobe of the CO flow shifts to the east corresponding to the
NH$_3$ wisp.  It has been suggested that molecular outflows associated
with jets may broaden over time, as has been seen in B5-IRS1; the
broadening flow could contribute to the eventual decline of accretion
(Velusamy \& Langer 1998).  Thus we may be seeing the HH 212 flow in
an early stage of broadening and sweeping out envelope material.

\subsection{Angular Momentum Release Through the Jet} 

Bipolar outflows appear to be a ubiquitous product of the formation of
stars.   Strong winds from young stellar sources are collimated into
jets which are observed in visible and infrared shock emission.
Magnetic fields may be responsible for channeling and collimating the
jets (see, e.g., Shu et al. 1995).

Outflows are thought to be a 'release valve' for the angular momentum
carried into the protostellar accretion region by infalling material.
Otherwise, matter infalling from even a very slowly rotating circumstellar
envelope will have too much conserved angular momentum to fall into
the inner accretion region, unless sufficient angular momentum is lost
through disk viscosity.  

{\em{Observational}} evidence for jets as significant angular momentum
carriers has been virtually nonexistent.  Recently, however, results
of slit spectroscopy on the shock knots in the HH 212 jet have given
some evidence of jet rotation (Davis et al. 2000).  Five of six knots
studied show similar evidence of rotation; in the inner part of the
southern flow, for example, a clear velocity gradient is seen.  If
this gradient represents rotation, {\em{the jet rotation is in the
same directional sense as that of the rotating molecular gas envelope
we observe in ammonia.}} A rough extrapolation of the large scale
angular momentum in the envelope gas fits the observed gradient in the
jet, assuming a fraction of infalling mass is channeled into the flow.
Thus this may be the first observational evidence for a rotating jet
carrying away the angular momentum of material falling in from a
rotating envelope.  Clearly more detailed observations and evidence of
jet rotation and inner disk kinematics are needed.


\acknowledgments
We thank our referee and also P. T. P. Ho for insightful guidance.
Support for J.W. was provided by NASA through Hubble Fellowship
grant \#HF-01115.01-98A awarded by the Space Telescope Science
Institute, which is operated by the Association of Universities for
Research in Astronomy, Inc., for NASA under contract NAS 5-26555.


\clearpage



\figcaption[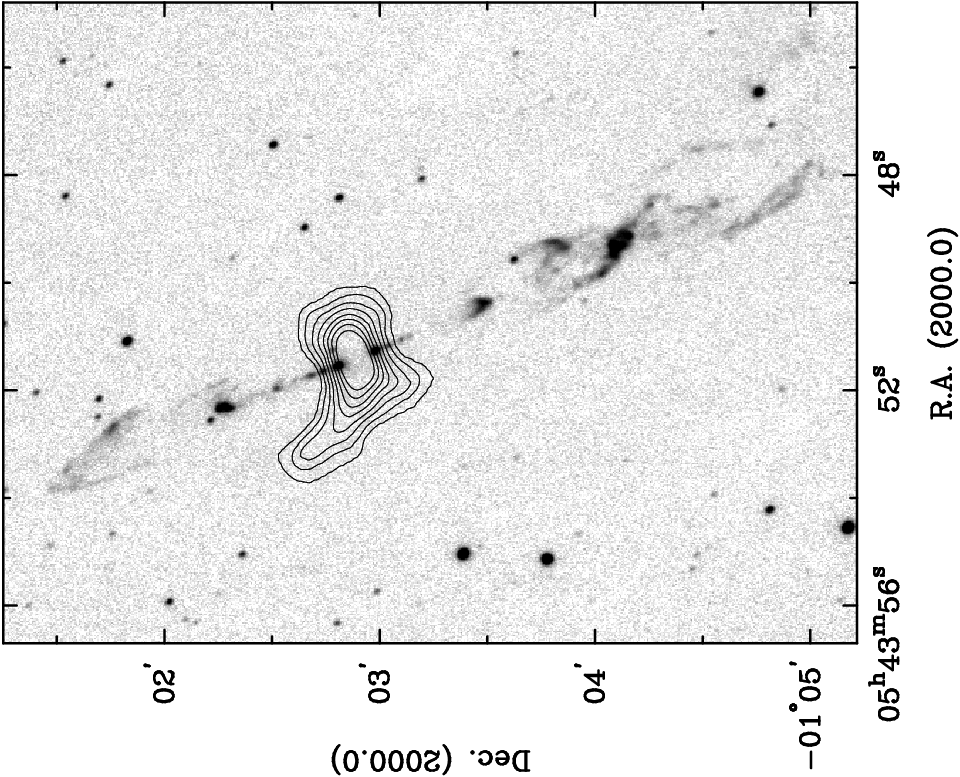]{
The HH~212 jet, as imaged in H$_2$ 2.12 micron shock emission
(Zinnecker, McCaughrean, \& Rayner 1998).  The observations
were carried out in Spain in 1997 with the Calar Alto 3.5-m telescope using its
Omega Prime wide field infrared camera and kindly given
courtesy of T. Stanke.  The contours represent
the integrated NH$_3$ (1,1) emission of the dense protostellar envelope
gas, as mapped with the VLA.  Contour levels begin at 
16 mJy beam$^{-1}$ km s$^{-1}$ and increase by steps of 
16  mJy beam$^{-1}$ km s$^{-1}$.
 \label{fig1}}

\figcaption[wiseman.fig2.ps]{(a): The velocity field of the dense gas
around the HH~212 jet, as imaged with the VLA in the ammonia (1,1)
line.  The shades represent the moment 1 velocity field, showing a
velocity gradient across the flattened gas envelope as evidence of
rotation.  Contours are the integrated intensity as in Figure 1.  The
arrows indicate the orientation of the HH 212 jet.  (b): Heating in
the HH 212 NH$_3$ core.  Contours are the integrated intensity of the
(1,1) line, as in Figure 1.  The shades represent the (2,2)/(1,1)
intensity ratio, which increases with temperature and  optical depth.
The inner region is slightly heated to 15 K, possibly where the
embedded YSO and jet are heating the surrounding gas. 
(c): Velocity-position cut across the FWHM long axis of the HH 212
flattened envelope, showing the gradient in first moment velocity.  
Positions are taken across the midline of
the flattened envelope of panel (a), with offsets taken from
the core center along an axis perpendicular to the jet. \label{fig2}}

\clearpage

\begin{figure}[!ht]
\begin{center}
\includegraphics[angle=270]{wiseman.fig1.ps}
\end{center}
\end{figure}
 
\clearpage
 
\begin{figure}[!ht]
\begin{center}
\plotone{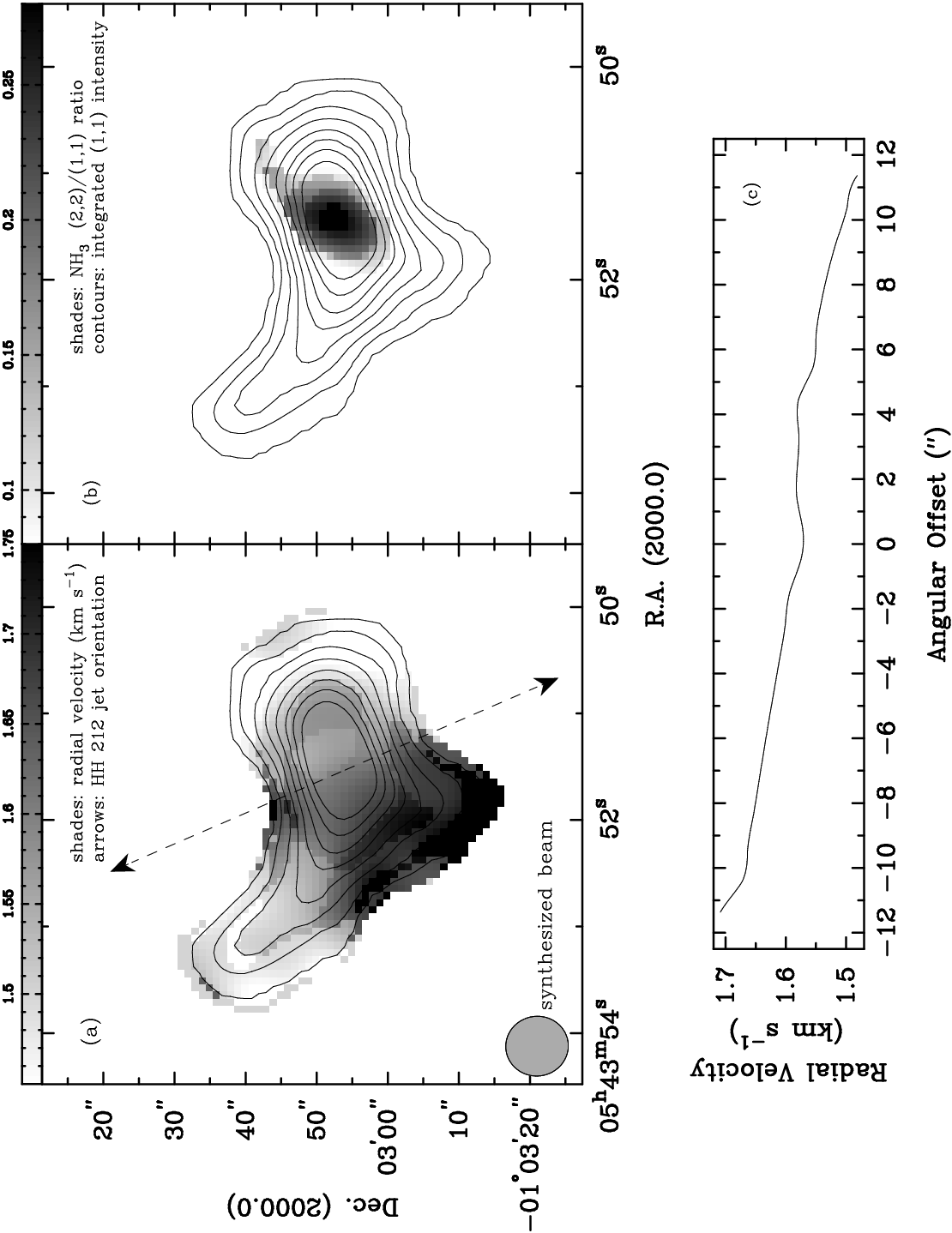}
\end{center}
\end{figure}
 
\clearpage
 
\begin{figure}[!ht]
\begin{center}
\plotone{wiseman.fig2.color.cps}
\end{center}
\end{figure}
 
\clearpage

\end{document}